\documentclass[preprint,12pt]{elsarticle}
\usepackage[a4paper,left=2.5cm,right=2.5cm,top=2.5cm,bottom=2.5cm]{geometry}
\biboptions{sort&compress}

\usepackage{amssymb}
\usepackage{amsmath}
\usepackage{booktabs}
\usepackage{tabularx}
\usepackage{makecell}
\usepackage{xcolor}
\usepackage{microtype}
\usepackage[font=small,labelfont=bf]{caption}
\usepackage{enumitem}
\setlist{itemsep=2pt,topsep=4pt}
\usepackage{titlesec}
\titlespacing*{\section}{0pt}{12pt}{6pt}
\titlespacing*{\subsection}{0pt}{9pt}{4pt}

\journal{Photonics and Nanostructures -- Fundamentals and Applications}

\begin{document}

\begin{frontmatter}

\title{Automated identification of quasi-BIC modes in disordered resonator chains using spatial field envelopes and resonance descriptors}

\author[msu,sber]{Vadim A. Porvatov}
\author[itmo]{Kliment V. Semushev}
\author[itmo]{Ravshanjon Kh. Nazarov}
\author[itmo]{Nikolai A. Vlasov}
\author[itmo]{Alexander I. Solomonov}
\author[itmo]{Zarina F. Kondratenko}
\author[itmo]{Ekaterina E. Maslova\corref{cor1}}
\ead{ekaterina.maslova@metalab.ifmo.ru}

\cortext[cor1]{Corresponding author}

\affiliation[msu]{
organization={Lomonosov Moscow State University},
addressline={Leninskie Gory, 1},
city={Moscow},
postcode={119991},
country={Russian Federation}
}

\affiliation[sber]{
organization={Sberbank},
addressline={Vavilova Street 19},
city={Moscow},
postcode={117312},
country={Russian Federation}
}

\affiliation[itmo]{
organization={School of Physics and Engineering, ITMO University},
addressline={49 Kronverksky Prospekt, Building A},
city={Saint Petersburg},
postcode={197101},
country={Russian Federation}
}

\begin{abstract}
Structural disorder turns bound states in the continuum (BICs) into quasi-BIC resonances and mixes them with spectrally close eigenmodes, so the target mode must be identified anew in every disorder realization. We develop an automated procedure for this task and test it on a chain of 50 ceramic disks, where disorder reorders the standing-wave eigenmodes at the edge of the BIC band: at disorder amplitude $\sigma=0.05$, the highest-quality-factor ($Q$) mode is the target in only 45\% of realizations, and above $\sigma\approx0.06$ the highest-frequency mode is no longer the target. The target is identified from the spatial envelope of its field and its resonance parameters by an ensemble of a convolutional--recurrent neural network and two gradient-boosted decision-tree models. On 299 held-out realizations, the ensemble selects the target in 82.6\% of cases, against 60.9\% and 49.5\% for the highest-frequency and highest-$Q$ rules, and in 69.5\% of cases for $\sigma\ge0.09$, where the best of these rules reaches 57.6\%. The mean $Q$ of the selected modes follows that of the manually labeled modes ($\langle Q\rangle\propto\sigma^{k}$, $k=-0.80\pm0.05$ versus $-0.86\pm0.04$). After adaptation on 50 labeled realizations, the workflow identifies the quasi-BIC of a Ge$_2$Sb$_2$Te$_5$ bar metasurface in 98--100\% of labeled cases, including a lossy phase where $Q$ does not distinguish the mode.
\end{abstract}


\end{frontmatter}

\section{Introduction}

Bound states in the continuum (BICs) in photonic structures provide a route to extremely high quality ($Q$) factor and strong field localization while remaining embedded in the radiation continuum \cite{Hsu2016_BIC_review,Koshelev2023-review,Zhang2026-review}. Such states have enabled a variety of applications, including highly sensitive sensors, nonlinear converters, and compact lasers \cite{Koshelev2018_asymmetricBIC,Kodigala2017_laser,Kang2023-applications}. Chains of subwavelength dielectric resonators offer a convenient platform for realizing and probing BICs, and symmetry-protected and accidental BICs have been observed in them experimentally and analyzed theoretically \cite{Sadrieva2019_chainBIC,Sidorenko2021_chainBIC,Bulgakov2019-finite,Nazarov2025_chain_disorder}. In these systems, structural disorder inevitably converts ideal BICs into quasi-BICs with finite radiative losses~\cite{Sadrieva2017-roughness,Ni2017-fluctuations,Maslova2021-disorder,Nazarov2025_chain_disorder,Centeno2025-emission}, modifying the spatial structure and localization properties of the eigenmodes~\cite{Segev2013-anderson,Lahini2008-anderson,Poddubny2012-fano,Maslova2021-disorder}.

In a disordered chain, each realization typically supports several eigenmodes clustered around the resonance frequency of the collective mode, and only one of them corresponds to the target BIC-like resonance~\cite{Nazarov2025_chain_disorder}. This target mode is usually identified by visual inspection of field profiles and their symmetries~\cite{Nazarov2025_chain_disorder}, as is also common in accelerating-cavity design~\cite{Georg2019-UQtracking,Ziegler2023-morphing}. Visual inspection is time-consuming and becomes impractical for large parameter sweeps or high-dimensional design spaces. Simple threshold-based criteria do not solve the problem, because the $Q$ factor of a single mode is broadly distributed with a long tail~\cite{Minkov2013-statistics}, and disorder strongly perturbs both the spatial profile and the $Q$ factor of the mode~\cite{Maslova2021-disorder,Nazarov2025_chain_disorder}.

Machine learning (ML) methods are increasingly used in nanophotonics~\cite{Ma2021-deep,Jiang2021-deepnn,ML_photonics_review} for tasks such as inverse design~\cite{Peurifoy2018-nanophotonic,Li2026-metadit}, band-structure prediction~\cite{Christensen2020-crystals,PhysicsInformedModes}, and mode classification~\cite{ML_photonics_review,Barth2018-modes,Ma2022-bic}. The closest works use ML to classify mode families in ordered structures~\cite{Ma2022-bic,Harsh2026-hybridmode}, to distinguish a specific class of eigenstates among many~\cite{Han2023-scardetector}, to resolve ambiguous inverse problems in resonant photonic systems~\cite{Liu2026-cmtnn}, and to classify eigenstates of disordered chains~\cite{Cybinski2025-ood}. Without ML, BICs have been identified among eigensolver modes through their sensitivity to boundary conditions~\cite{Laude2026-boundary}. Most existing approaches either operate on engineered scalar descriptors~\cite{Molokeev2023-forest} or focus on ideal periodic systems with limited structural variation~\cite{Ma2022-bic,Christensen2020-crystals}. Nevertheless, the problem of automatically identifying a specific physical eigenmode in a strongly disordered resonant system, based simultaneously on detailed field information and spectral characteristics, remains largely unexplored. Mode-tracking strategies for parametrized eigenvalue problems have been discussed in computational electromagnetics~\cite{Jorkowski2019modetracking,Raines2012-CMtracking,Georg2019-UQtracking,Ziegler2023-morphing,Li2026-MLCMtracking}, but they follow a mode along a parameter and are therefore not applicable to independent disorder realizations.

In this work, we address this gap by formulating the identification of a target magnetic-quadrupole mode in a disordered chain of dielectric disks as a binary classification problem. We generate a large dataset of eigenmodes from full-wave simulations of a chain of 50 dielectric disks with axial positional disorder, spanning a range of disorder amplitudes with many realizations per amplitude. For each eigenmode, we construct spatial envelopes of the electric-field modulus and of the real and imaginary parts of the field along the chain, and complement them with scalar metadata, namely the disorder amplitude, $Q$ factor, and complex eigenfrequency. Using this representation, we develop an ensemble classifier that combines a convolutional--recurrent neural network (CNN--BiGRU) operating on the field-modulus envelope with two gradient-boosted decision-tree models operating on envelope descriptors and scalar features, one of which also uses descriptors of each candidate relative to the other modes of its realization. The ensemble selects the target mode in 82.6\% of the held-out realizations despite a class imbalance of about 1:10 at the candidate level; its advantage over frequency- and $Q$-based selection rules is largest at strong disorder, where these rules fail. We further show that, after adaptation on 5\% of the data, the workflow transfers to a Ge$_2$Sb$_2$Te$_5$ (GST) bar metasurface, a phase-change platform~\cite{Wuttig2017-pcm,Bochek2022-gst}, and recovers the expected $Q(\sigma)$ scaling. For the chain, the mean total $Q$ factor of the selected modes reproduces that of the manually labeled modes.

\section{Disordered resonator chain and eigenmode dataset}
\label{sec:problem}

\subsection{Photonic system}
\label{subsec:photonic-system}

We consider a chain of $N=50$ dielectric disks with a period $L = 34$~mm, supporting a magnetic-quadrupole (MQ) quasi-BIC mode. The mode originates from a symmetry-protected BIC at the $\Gamma$-point~\cite{Nazarov2025_chain_disorder}. Each disk has a relative permittivity $\varepsilon = 44$, a loss tangent $\tan\delta = 10^{-4}$, a radius $R = 15$~mm, and a height $h = 20$~mm. The chain is described in detail in Ref.~\cite{Nazarov2025_chain_disorder}. Structural disorder is introduced through independent axial displacements of the disk centers: the $i$-th disk is displaced from its lattice site $z_i^{0}=iL$ by
\begin{equation}
\delta z_i =\xi_i \sigma L,
\label{eq:disorder}
\end{equation}
where $\xi_i$ are independent random variables uniformly distributed on $[-1, 1]$, and $\sigma$ is the dimensionless disorder amplitude (in units of the period). All disk centers remain on the common chain axis; disorder acts exclusively through irregular axial spacing between neighboring disks.

\subsection{Mode ladder and loss budget}
\label{subsec:ladder}

\begin{figure*}[t]
\centering
\includegraphics[width=1\linewidth]{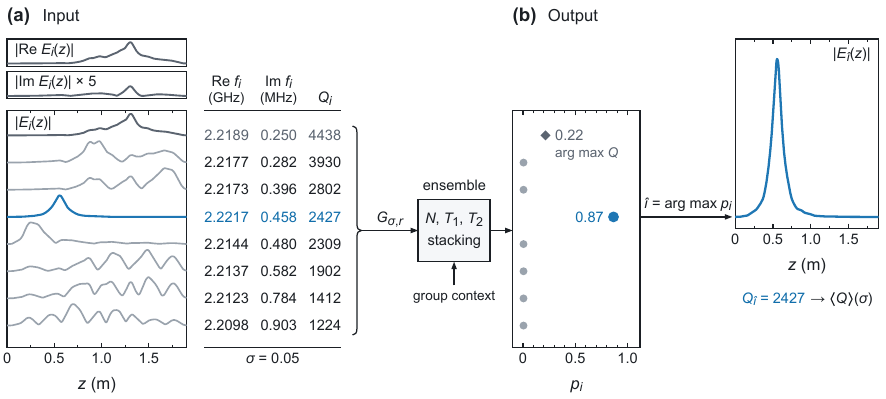}
\caption{Target-mode identification in a single disorder realization of the chain ($\sigma=0.05$, test split). (a)~Input: the eight candidate eigenmodes of one solution group $G_{\sigma,r}$, one per row and ordered by $Q$; for each candidate, the envelope $|E_i(z)|$, the real and imaginary parts of its complex eigenfrequency, and $Q_i$. The top strips show the unsigned envelopes $|\mathrm{Re}\,E_i|$ and $|\mathrm{Im}\,E_i|$ (the latter magnified five times) of the top-row candidate. All envelopes are divided by the peak of $|E_i|$ and, for display only, averaged over one disk period; the classifier reads the unaveraged envelopes. Label-free group context, i.e.\ the parameters of a candidate relative to those of the other candidates, is computed from the whole group. The ensemble (Section~\ref{sec:method}, Fig.~\ref{fig:framework}) combines the scores of the neural branch $N$ and the tree branches $T_1$ and $T_2$. (b)~Output: the probability $p_i$ that candidate $i$ is the target; the candidate $\widehat{i}$ with the largest $p_i$ is selected (blue; $p_i=0.87$, $Q=2427$; it coincides with the manual label), whereas the highest-$Q$ candidate (dark grey, $Q=4438$) scores $0.22$. The $Q$ factors of the selected modes, averaged over realizations, give $\langle Q\rangle(\sigma)$ (Fig.~\ref{fig:figure3}). In this realization, the target is also the highest-frequency candidate.}
\label{fig:figure1}
\end{figure*}

In the ordered chain ($\sigma=0$), the MQ band hosts a ladder of longitudinal standing waves (Fabry--P\'erot-like resonances of the finite chain). The target mode is the fundamental, single-lobe harmonic $n=1$ at the upper edge of the band, and the neighboring candidates are the harmonics $n=2,3,\ldots$ of the same band. Even without disorder, this mode is a quasi-BIC rather than an ideal BIC: its $Q$ factor is limited by the finite chain length and by absorption, $Q_0\approx8.9\times10^{3}$, which is close to the material limit $1/\tan\delta=10^{4}$. Disorder adds a third loss channel, so that
\begin{equation}
\frac{1}{Q}=\frac{1}{Q_N}+\frac{1}{Q_{\mathrm{abs}}}+\frac{1}{Q_{\mathrm{dis}}(\sigma)},
\label{eq:loss-budget}
\end{equation}
as in Ref.~\cite{Nazarov2025_chain_disorder}. Throughout this work, $Q$ denotes the total $Q$ factor returned by the eigensolver, which contains all three contributions; we do not separate them. Disorder shifts the target mode toward higher frequencies and mixes it with the higher harmonics and with defect modes localized on a few disks. As a result, neither the eigenfrequency nor the $Q$ factor alone provides a sufficiently robust criterion for identifying the mode of interest over the entire disorder range. Figure~\ref{fig:figure1} illustrates such a case; without an automated selection procedure, such cases must be resolved manually, one realization at a time.

\subsection{Eigenmode dataset and labeling}
\label{subsec:chain-dataset}

The eigenmodes of the $50$-disk chain were computed with the finite element method, focusing on the MQ quasi-BIC mode. The disorder amplitude was varied from $\sigma=0.01$ to $0.15$ in steps of $0.01$ (the ordered chain is not part of the dataset), with $99$--$140$ independent realizations per value. For larger $\sigma$, where the target mode cannot be identified in every realization, the number of realizations was increased so that $99$--$103$ realizations with an identified target are available for each value of $\sigma$.
Each eigenmode calculation yielded $8$, $10$, or $12$ modes in the vicinity of the collective resonance frequency of the chain ($\sim 2.22$~GHz). The electric-field profiles $E(z)$ (here and throughout the manuscript for brevity we denote as $E$ either the angular component $E_\varphi$ of the electric field (for the ceramic disk chain) or the component of magnetic field parallel to bars' own axis $H_y$ (for the GST bars array)) were sampled along the chain axis on a non-uniform grid. The candidates include the target quasi-BIC mode, higher harmonics of the same band, and disorder-induced states localized on a few disks.

Every candidate was stored together with its complex eigenfrequency, $Q$ factor, and disorder amplitude. For $\sigma\le0.05$, the target was taken to be the highest-frequency mode of the MQ band, which is the band-edge standing wave of the ordered chain (Section~\ref{subsec:ladder}). For $\sigma\ge0.06$, the target was identified manually following Ref.~\cite{Nazarov2025_chain_disorder}: candidates were compared with the mean target frequency at the preceding $\sigma$, and modes localized on two or three neighboring disks were excluded in favor of modes extended along the chain. This procedure produced a dataset of $15{,}784$ candidate modes grouped into $1{,}707$ independent eigensolution groups. Among them, $1{,}503$ groups contain exactly one positively labeled mode, and $204$ contain no positive candidate because the target could not be identified unambiguously (all at $\sigma\geq0.06$). Groups without a positive label are retained during candidate-level training but excluded from the group-level top-1 evaluation. The resulting positive-to-negative class ratio at the candidate level is approximately $1\!:\!10$.

The dataset was divided into training, validation, and test parts in the proportion $60\%/20\%/20\%$ using group-wise random splitting with seed $1{,}648$, so that all candidates from a given realization belong to the same part. The resulting parts contain $9{,}442$, $3{,}142$, and $3{,}200$ candidate modes, respectively; the test part contains $342$ groups, $299$ of them with an identified target. The test set was not used for feature standardization, hyperparameter selection, stacking, or threshold selection.

\subsection{Formal problem definition}
\label{subsec:problem-formulation}

For each disorder amplitude $\sigma$ and disorder realization $r$, the eigensolver produces a group $G_{\sigma,r}$ containing several candidate modes. A candidate $i\in G_{\sigma,r}$ is represented by its one-dimensional field profile and a vector of scalar characteristics:
\begin{equation}
\mathcal{X}_i = \left\{E_i(z),\mathbf{m}_i\right\},
\qquad
\mathbf{m}_i =
\left[
\sigma_i,
Q_i,
\mathrm{Re}\,f_i,
\mathrm{Im}\,f_i
\right],
\label{eq:candidate-input}
\end{equation}
where $Q_i$ is the total $Q$ factor (including material absorption), and $\mathrm{Re}\,f_i$ and $\mathrm{Im}\,f_i$ denote the real and imaginary parts of the complex eigenfrequency $f_i$ (the COMSOL variable \texttt{emw.freq}, in Hz). The vector $\mathbf{m}_i$ contains three independent quantities, since $Q_i=\mathrm{Re}\,f_i/(2|\mathrm{Im}\,f_i|)$. Frequencies are quoted in GHz. The binary label $y_i\in\{0,1\}$ indicates whether the candidate is the physical mode of interest.

The model assigns a probability $p_i=P(y_i=1\mid\mathcal{X}_i,G_{\sigma,r})$ to each candidate; it depends on the other candidates of the group through the group-relative features of Section~\ref{subsec:spectral-representation}. The final prediction for a group is obtained by selecting the candidate with the highest probability:
\begin{equation}
\widehat{i}_{\sigma,r}
= \operatorname*{arg\,max}_{i\in G_{\sigma,r}} p_i.
\label{eq:group-selection}
\end{equation}
Hence, the learning problem is formulated at the candidate level, whereas the physical decision (selecting one target mode among several spectrally close candidates; Fig.~\ref{fig:figure1}) is made at the level of complete groups. Because the argmax always returns a candidate, the selector by itself does not detect realizations without an identifiable target; such groups are excluded from the group-level evaluation (Section~\ref{subsec:limitations}).

\section{Field- and resonance-aware mode-identification framework}
\label{sec:method}

\begin{figure}
\centering
\includegraphics[width=1\linewidth]{
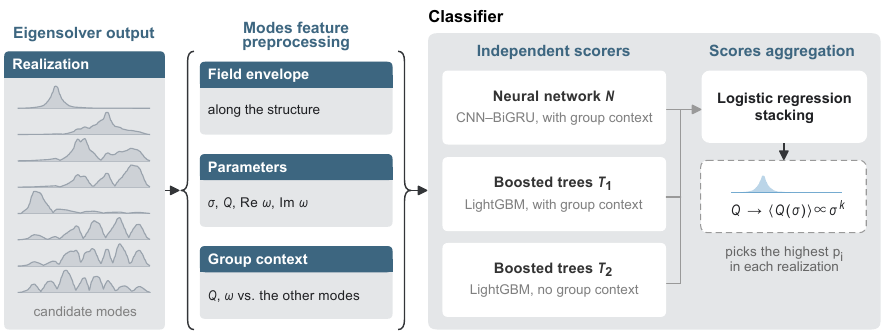}
\caption{Mode-identification framework. For disorder amplitude $\sigma$ and realization $r$, the eigensolver returns a group $G_{\sigma,r}$ of candidate modes. Each candidate is described by its field envelope, its resonance parameters $\mathbf{m}_i$, and its position within the group (group context). Three independent scorers are trained on the labeled groups of the training split: the CNN--BiGRU network $N$ reads the modulus envelope $A_i(z)$, conditioned on $\mathbf{m}_i$ and part of the group context; the LightGBM models $T_1$ and $T_2$ use descriptors of the $|\mathrm{Re}\,E_i|$, $|\mathrm{Im}\,E_i|$ envelopes together with $\mathbf{m}_i$, and $T_1$ also uses the group context. Logistic stacking combines the three seed-averaged scores into the probability $p_i$ that candidate $i$ is the target, and the highest-scoring candidate of each group is selected [Eq.~\eqref{eq:group-selection}]. The $Q$ factors of the selected modes give $\langle Q\rangle(\sigma)$.}
\label{fig:framework}
\end{figure}

This section specifies the model input (Fig.~\ref{fig:framework}), describes the three branches that process it and the fusion of their outputs, and gives the training configuration.

\subsection{Model input}
\label{subsec:model-input}

The raw simulation data were converted into a common array format (NumPy NPZ). Only the field profile along the inner line cut (at $r=7.5$~mm, half the disk radius) was retained. Before export, the field profile of every candidate was divided by its peak magnitude; the corresponding normalization coefficient was stored as $\mathrm{norm}_{\max}$. This transformation removes the arbitrary global scale of the numerical eigenvector while retaining its spatial structure. Here, $\mathrm{norm}_{\max}$ is the peak of $|E(z)|$ along this line in the units of the eigensolver output (V/m); it therefore inherits the solver's normalization of the eigenvector. No phase normalization was applied: the real and imaginary parts are taken as returned by the eigensolver.

The raw field profiles contain rapid subwavelength oscillations associated with the periodic disk geometry. We represent each profile by its upper spatial envelope, extracted separately for $|E(z)|$, $\mathrm{Re}\,E(z)$, and $\mathrm{Im}\,E(z)$ by linear interpolation between successive local maxima. No smoothing was applied, so as to preserve secondary maxima of the field profile. The envelope profiles were resampled onto a common uniform grid with a spacing $\Delta z \approx 1$~mm and $L' = 1883$ points ($z=0$--$1.9$~m), covering the full length of the chain. The line cut starts at $z=50$~mm, so the first 50 points are zero padding.

The envelopes of the real and imaginary parts are stored as unsigned envelopes, $|\mathrm{Re}\,E_i|$ and $|\mathrm{Im}\,E_i|$ (the sign and phase of the field are therefore not retained), and form, for each candidate, the two-channel sequence
\begin{equation}
\mathbf{S}_i =
\left[
\left|\mathrm{Re}\,E_i(z_1)\right|,
\left|\mathrm{Im}\,E_i(z_1)\right|;
\ldots;
\left|\mathrm{Re}\,E_i(z_{L'})\right|,
\left|\mathrm{Im}\,E_i(z_{L'})\right|
\right]
\in\mathbb{R}^{L'\times 2}.
\label{eq:field-sequence}
\end{equation}
The envelope of the field modulus, $A_i(z)$, obtained from $|E_i(z)|$ in the same way, is retained as an additional single-channel input. In summary, the model receives the following inputs for each candidate: (i) the standardized modulus envelope $A_i(z)$, processed by the neural branch (Section~\ref{subsec:neural-encoder}); (ii) the two-channel envelope sequence $\mathbf{S}_i$, processed by the tree-based branches (Sections~\ref{subsec:spectral-representation} and~\ref{subsec:tree-encoder}); (iii) the four-dimensional metadata vector $\mathbf{m}_i$ defined in Eq.~\eqref{eq:candidate-input}, used by all branches; and (iv) group-relative descriptors that characterize the candidate with respect to the other members of its group (Section~\ref{subsec:spectral-representation}).

Candidate modes are assigned to groups using the pair $(\sigma,r)$. The dataset is split by whole groups rather than by individual candidates, so that all modes from the same realization fall into the same split. Both the field channels and the four metadata variables are standardized using statistics computed on the training set only:
\begin{equation}
\widetilde{x}
= \frac{x-\mu_{\mathrm{train}}}{s_{\mathrm{train}}},
\label{eq:standardization}
\end{equation}
where $\mu_{\mathrm{train}}$ and $s_{\mathrm{train}}$ are the training-set mean and standard deviation, respectively. The same transformations are applied without refitting to the validation, test, and external-domain datasets.

\subsection{Envelope descriptors for the tree-based branches}
\label{subsec:spectral-representation}

The tree-based branches transform every standardized two-channel sequence $\widetilde{\mathbf{S}}_i$ into a fixed-dimensional representation. For each of the $|\mathrm{Re}\,E|$ and $|\mathrm{Im}\,E|$ channels, we compute statistical, spectral, shape, and autocorrelation descriptors. Here, ``spectral'' refers to the spatial Fourier transform of the envelope, not to the eigenfrequency spectrum.

The statistical block contains the mean, standard deviation, skewness, kurtosis, minimum, maximum, range, median, interquartile range, and root mean square. The spectral block is obtained from the discrete Fourier transform and includes the frequencies and magnitudes of the five dominant components, the spectral centroid, spectral entropy, and the relative energies of the low-, middle-, and high-frequency bands. The shape block contains the number of mean-level crossings, the number and maximum amplitude of local peaks, and the mean and standard deviation of the distances between consecutive peaks. Finally, the autocorrelation function is evaluated at lags of 1, 5, 10, and 50 grid points.

The channel-wise descriptors are extended by six statistics (mean, standard deviation, minimum, maximum, range, and root mean square) of the pointwise norm of the two standardized channels and by four descriptors of the channel-angle quantity
\begin{equation}
\phi_i(z)
= \operatorname{atan2}
\left(
\widetilde{E}_i^{\mathrm{Im}}(z),
\widetilde{E}_i^{\mathrm{Re}}(z)
\right),
\label{eq:channel-angle}
\end{equation}
where $\widetilde{E}_i^{\mathrm{Re}}$ and $\widetilde{E}_i^{\mathrm{Im}}$ are the two standardized channels of $\widetilde{\mathbf{S}}_i$. Because they are computed from standardized unsigned envelopes, this norm and $\phi_i$ characterize the relative contribution of the two channels and should not be interpreted as the modulus or the physical phase of the complex field. The resulting field representation contains $78$ features. After concatenation with the four standardized metadata variables, the candidate feature vector is
\begin{equation}
\mathbf{z}_i
= \operatorname{CONCAT}
\left(
\widetilde{\mathbf{m}}_i,
\Phi(\widetilde{\mathbf{S}}_i)
\right)
\in\mathbb{R}^{82},
\label{eq:feature-vector}
\end{equation}
where $\Phi(\cdot)$ denotes the complete field-feature extractor. Non-finite feature values are replaced by zeros after extraction.

To account for the position of a candidate within its group, we additionally compute group-relative descriptors. For $Q$, $\mathrm{Re}\,f$, $\mathrm{Im}\,f$, and $\mathrm{norm}_{\max}$, we compute the descending rank within the group, the within-group standardized value, and the ratio to the group maximum; the group size is appended as an additional scalar. These transformations do not use target labels and produce $13$ additional features $\mathbf{g}_i$, which form the extended input
\begin{equation}
\mathbf{z}_i^{+}
= \operatorname{CONCAT}\left(\mathbf{z}_i, \mathbf{g}_i\right)
\in\mathbb{R}^{95}.
\label{eq:feature-vector-extended}
\end{equation}

\subsection{Tree-based branches (LightGBM)}
\label{subsec:tree-encoder}

Two LightGBM classifiers, gradient-boosted decision-tree models, process the two tree inputs. The first branch, $T_1$, operates on the extended vector $\mathbf{z}_i^{+}$; the second branch, $T_2$, operates on $\mathbf{z}_i$ only. The output of each branch is a sum of decision trees:
\begin{equation}
F_{T_k}(\mathbf{x}_i)
= \sum_{b=1}^{B}\eta\, T_b^{(k)}(\mathbf{x}_i),
\qquad
p_i^{T_k}
= \operatorname{sigmoid}
\left(F_{T_k}(\mathbf{x}_i)\right),
\qquad k=1,2,
\label{eq:tree-encoder}
\end{equation}
where $\mathbf{x}_i=\mathbf{z}_i^{+}$ for $k=1$ and $\mathbf{x}_i=\mathbf{z}_i$ for $k=2$, $T_b^{(k)}$ is the $b$-th decision tree, $B$ is the number of boosting iterations, and $\eta$ denotes the learning rate. The branch $T_1$ combines the global properties of an individual candidate with its characteristics relative to the other candidates in its group. The branch $T_2$ has no access to the group-relative block and estimates the probability from the field morphology and the absolute resonance parameters alone.

\subsection{Neural branch (CNN--BiGRU)}
\label{subsec:neural-encoder}

The neural branch operates directly on the standardized modulus envelope $A_i(z)$. Its first part consists of three one-dimensional convolutional blocks. Each block consists of a convolution, batch normalization, ReLU activation, and max pooling with a factor of two. The blocks have $16$, $32$, and $64$ output channels and kernel sizes of $7$, $5$, and $3$, respectively.

The resulting sequence of local field embeddings, of length $T\approx L'/8$, is processed by a three-layer bidirectional gated recurrent unit (BiGRU). Each direction has a hidden dimension of $125$, producing the sequence
\begin{equation}
\mathbf{H}_i
= \left[
\mathbf{h}_{i,1},\ldots,\mathbf{h}_{i,T}
\right],
\qquad
\mathbf{h}_{i,t}\in\mathbb{R}^{250}.
\label{eq:bigru-output}
\end{equation}
To aggregate this sequence, we apply learnable attention pooling~\cite{Bahdanau2015-attention,Yang2016-HAN}:
\begin{equation}
a_{i,t}
= \frac{
\exp\left(\mathbf{q}^{\mathsf{T}}\mathbf{h}_{i,t}\right)
}{
\sum_{j=1}^{T}
\exp\left(\mathbf{q}^{\mathsf{T}}\mathbf{h}_{i,j}\right)
},
\qquad
\mathbf{s}_i
= \sum_{t=1}^{T}a_{i,t}\mathbf{h}_{i,t},
\label{eq:attention-pooling}
\end{equation}
where $\mathbf{q}$ is a trainable attention vector.

The scalar information is incorporated via feature-wise linear modulation (FiLM). The conditioning vector $\widetilde{\mathbf{c}}_i\in\mathbb{R}^{11}$ consists of the four standardized metadata variables $\widetilde{\mathbf{m}}_i$ and seven group-relative values (the descending rank, the within-group standardized value, and the ratio to the group maximum of $Q$ and $\mathrm{Im}\,f$, and the group size), standardized with training-set statistics. A linear projection of $\widetilde{\mathbf{c}}_i$ produces the vectors $\boldsymbol{\gamma}_i$ and $\boldsymbol{\beta}_i$, which modulate the pooled field representation:
\begin{equation}
\left[
\boldsymbol{\gamma}_i,
\boldsymbol{\beta}_i
\right]
= W_{\mathrm{M}}\widetilde{\mathbf{c}}_i+\mathbf{b}_{\mathrm{M}},
\qquad
\widetilde{\mathbf{s}}_i
= \boldsymbol{\gamma}_i\odot\mathbf{s}_i
+ \boldsymbol{\beta}_i.
\label{eq:film-conditioning}
\end{equation}
The resulting vector is passed through dropout, a linear layer with SiLU activation, and the final scalar projection:
\begin{equation}
p_{i}^{\mathrm{N}}
= \operatorname{sigmoid}
\left[
W_2\,
\operatorname{SiLU}
\left(
W_1\operatorname{Dropout}
(\widetilde{\mathbf{s}}_i)
\right)
\right]
\equiv \operatorname{sigmoid}(\zeta_i).
\label{eq:neural-output}
\end{equation}
The neural branch is trained with a hybrid objective that combines the candidate-level classification loss with a group-level ranking term:
\begin{equation}
\mathcal{L}
= \mathcal{L}_{\mathrm{BCE}}^{w}
+ \frac{1}{|\mathcal{G}_{B}|}\sum_{G\in\mathcal{G}_{B}}
\left[
-\log\frac{\exp\left(\zeta_{i^{*}(G)}\right)}
{\sum_{j\in G}\exp\left(\zeta_{j}\right)}
\right],
\label{eq:hybrid-loss}
\end{equation}
where $\mathcal{L}_{\mathrm{BCE}}^{w}$ is the binary cross-entropy with the positive-class weight equal to the ratio of negative to positive training candidates, $\zeta_j$ are the logits of Eq.~\eqref{eq:neural-output}, $\mathcal{G}_{B}$ is the set of groups in the mini-batch that contain exactly one positive candidate, and $i^{*}(G)$ is that candidate. The second term is the cross-entropy of selecting the target within its group and therefore directly optimizes the group-level decision of Eq.~\eqref{eq:group-selection}. Mini-batches are formed by packing complete groups until at least $64$ candidates are collected; groups without a positive label contribute to the binary term only.

\subsection{Probability fusion and group-wise selection}
\label{subsec:probability-fusion}

Each branch is trained with three random seeds ($42$, $133$, and $777$), and its output is the arithmetic mean of the three predicted probabilities:
\begin{equation}
\overline{p}_i^{\,b}
= \frac{1}{3}\sum_{s=1}^{3}p_{i,s}^{\,b},
\qquad b\in\{T_1,T_2,\mathrm{N}\}.
\label{eq:seed-averaging}
\end{equation}
The three branches are combined by a logistic stacking layer:
\begin{equation}
p_i
= \operatorname{sigmoid}
\left(
w_{1}\overline{p}_i^{\,T_1}
+ w_{2}\overline{p}_i^{\,T_2}
+ w_{3}\overline{p}_i^{\,\mathrm{N}}
+ b
\right).
\label{eq:probability-fusion}
\end{equation}
The stacking coefficients are fitted with $L_2$ regularization using only the validation predictions: $w_1=2.87$, $w_2=0.99$, $w_3=4.46$, and $b=-5.35$. For candidate-level classification, the decision threshold $t$ is selected on the validation set by maximizing the F1 score over the 201 quantiles of the validation scores ($t=0.379$). The group-level selection of Eq.~\eqref{eq:group-selection} depends only on the relative ordering of $p_i$ and does not require a threshold.

\subsection{Training configuration and evaluation metrics}
\label{subsec:training-configuration}

Both tree branches use the same LightGBM configuration: $870$ boosting iterations with a learning rate of $5.78\times10^{-3}$, a maximum tree depth of $10$, at most $110$ leaves, at least $69$ samples per leaf, and a feature fraction of $0.567$ per tree, trained deterministically with column-wise histogram construction. This configuration was obtained by a $50$-trial search with the tree-structured Parzen estimator (TPE), a Bayesian hyperparameter-optimization algorithm implemented in the Optuna library, on an earlier export of the chain dataset; on the validation split it was preferred over the configuration of a later $60$-trial search ($1{,}001$ iterations, learning rate $1.164\times10^{-2}$, depth $4$, $64$ leaves) by the group-level top-1 accuracy ($0.812$ versus $0.805$).

The CNN--BiGRU architecture was selected from $185$ completed Optuna TPE trials, each trained for at most $60$ epochs with early stopping after $12$ epochs without an improvement in the validation loss. The final CNN--BiGRU models were trained with the Adam optimizer at a learning rate of $4.24\times10^{-4}$, a dropout probability of $0.171$, and the objective of Eq.~\eqref{eq:hybrid-loss} for at most $120$ epochs. After each epoch, the group-level top-1 accuracy on the validation split was evaluated; the learning rate was reduced by a factor of $0.25$ after $8$ epochs without its improvement, training stopped after $15$ such epochs, and the epoch with the highest validation accuracy (ties broken by the validation ranking loss) was retained. All neural computations used deterministic GPU kernels.

Model performance was assessed with two complementary sets of metrics. The primary, physically meaningful measure is the group-level top-1 accuracy: the fraction of labeled groups for which the highest-scoring candidate, selected via Eq.~\eqref{eq:group-selection}, is the target mode; it is reported with exact (Clopper--Pearson) $95\%$ confidence intervals. As complementary candidate-level measures, we used the F1 score, the area under the receiver-operating-characteristic curve (ROC--AUC), and the Matthews correlation coefficient (MCC) at the validation-selected threshold. The model is compared with simple selection rules based on the eigenfrequency and the $Q$ factor (Section~\ref{subsec:simple-rules}); paired group-level outcomes are compared with the exact McNemar test.

The experiments were conducted on a workstation equipped with an AMD Ryzen 9 9950X processor, $64$~GB of RAM, and an NVIDIA GeForce RTX 5090 GPU with $32$~GB of video memory. The implementation used Python 3.11, LightGBM 4.7, PyTorch 2.11, CUDA 12.8, and Optuna 4.9.

\section{Results and discussion}
\label{sec:results}

\subsection{Where frequency and $Q$ fail}
\label{subsec:simple-rules}

Before evaluating the classifier, we determine where simple selection rules stop working. Table~\ref{tab:simple_rules} gives, for the $299$ labeled test groups, the fraction of groups in which each rule selects the labeled target. Over all $1{,}503$ labeled groups, the target is the highest-frequency mode of its group for every $\sigma\le0.06$; note that for $\sigma\le0.05$ this follows from the labeling rule of Section~\ref{subsec:chain-dataset}. The rule based on the maximum $Q$ fails earlier: it selects the target in $97\%$ of the groups at $\sigma=0.03$, $72\%$ at $\sigma=0.04$, and $45\%$ at $\sigma=0.05$. At $\sigma\approx0.07$--$0.09$, the regime changes: additional modes appear above the target in frequency, the groups grow from 8 to 10--12 modes, and the frequency rule drops to $81\%$ ($\sigma=0.07$), $75\%$ ($\sigma=0.08$), and $18$--$39\%$ for $\sigma\ge0.09$. The modes above the target are strongly localized: for $\sigma\ge0.07$, the median participation ratio of their envelopes, $\mathrm{PR}=(\int|E|^2dz)^2/(L_{\mathrm{ch}}\int|E|^4dz)$, is $0.07$, compared with $0.11$ for the target, which corresponds to a few disks. This is consistent with the defect modes described for longer chains in Ref.~\cite{Nazarov2025_chain_disorder}. In this regime, neither frequency nor $Q$ identifies the target, and field information is needed. Combining the two criteria, i.e., selecting the highest-frequency mode among the two or three highest-$Q$ modes, helps at moderate disorder but still selects the target in only $55$--$58\%$ of the groups with $\sigma\ge0.09$. A random choice within a group gives about $1/(\text{number of modes})\approx8$--$12\%$.

\begin{table}[t]
\centering
\caption{Fraction of the $299$ labeled test groups in which a simple rule selects the labeled target mode, by disorder amplitude $\sigma$ (numbers of groups in parentheses).}
\label{tab:simple_rules}
\small
\renewcommand{\arraystretch}{1.25}
\setlength{\tabcolsep}{6pt}
\begin{tabularx}{\textwidth}{@{}>{\raggedright\arraybackslash}X *{4}{c}@{}}
\toprule
 & \multicolumn{3}{c}{\textbf{Range of $\sigma$}} & \\
\cmidrule(lr){2-4}
\textbf{Selection rule} & $\le 0.05$ (87) & $0.06$--$0.08$ (61) & $\ge 0.09$ (151) & \textbf{All} (299) \\
\midrule
Highest frequency & 1.000 & 0.852 & 0.285 & 0.609 \\
Highest $Q$ & 0.839 & 0.180 & 0.424 & 0.495 \\
Highest frequency of the 2 highest-$Q$ modes & 0.966 & 0.705 & 0.576 & 0.716 \\
Highest frequency of the 3 highest-$Q$ modes & 0.989 & 0.885 & 0.550 & 0.746 \\
\midrule
Random choice & 0.125 & 0.121 & 0.096 & 0.109 \\
\bottomrule
\end{tabularx}
\end{table}

\subsection{Identification accuracy and ablations}
\label{subsec:chain-performance}

\begin{figure}[h!]
\centering
\includegraphics[width=1\linewidth]{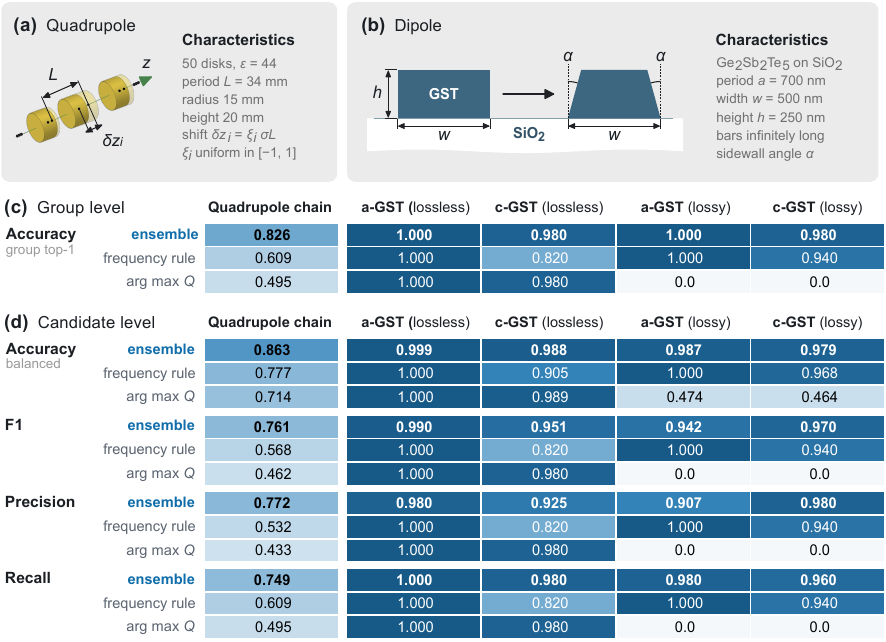}
\caption{Physical platforms and identification performance. (a)~Ceramic disk chain (MQ quasi-BIC) and (b)~GST bars on SiO$_2$ (dipole-like quasi-BIC). Each column of (c) and (d) is one dataset: the chain test split (299 labeled groups) and the four GST datasets (50 labeled groups each; the ensemble is adapted on the other labeled groups and scored out of fold). Every block compares the same three methods: the ensemble, the frequency rule (the candidate at a fixed frequency rank, fitted on other groups; for the chain, the highest-frequency candidate), and $\arg\max Q$; cell color encodes the value. (c)~Group level: each method picks one candidate per group; top-1 accuracy is the fraction of labeled groups in which the pick is the manually labeled target (exact $95\%$ confidence intervals: chain $[0.778, 0.867]$ for the ensemble, $[0.551, 0.664]$ for the frequency rule, $[0.437, 0.553]$ for $\arg\max Q$; GST $[0.929, 1]$ for 50/50, $[0.894, 0.999]$ for 49/50, $[0, 0.071]$ for 0/50). (d)~Candidate level: every candidate is classified as target or not; the ensemble flags candidates with $p_i$ above an F1-optimal threshold chosen on other data, and a rule flags one candidate per group. Balanced accuracy, the mean of the fractions of correctly classified targets and non-targets, is insensitive to the 1:10 (chain) and 1:20 (GST) class imbalance; a classifier that flags nothing scores 0.5, whereas its plain accuracy would be 0.91--0.95.}
\label{fig:figure4}
\end{figure}

Table~\ref{tab:performance} and Fig.~\ref{fig:figure4} report the performance of the models on the chain test set. The ensemble classifier identifies the target MQ quasi-BIC mode in $247$ of the $299$ labeled test groups (group-level top-1 accuracy $0.826$, $95\%$ confidence interval $[0.778, 0.867]$), compared with $0.609$ $[0.551, 0.664]$ for the highest-frequency rule and $0.495$ $[0.437, 0.553]$ for the highest-$Q$ rule (exact McNemar test, $p<10^{-11}$ for both). Because the labels for $\sigma\le0.05$ come from a rule rather than from the field, the accuracy is reported separately for the rule-labeled ($\sigma\le0.05$) and the manually labeled ($\sigma\ge0.06$) ranges. For $\sigma\ge0.09$, the ensemble identifies the target in $105$ of $151$ groups ($0.695$), compared with $0.576$ for the best rule in this range, the highest-frequency mode among the two highest-$Q$ modes (Table~\ref{tab:simple_rules}). In $36$ of the $299$ test groups, both the highest-frequency and the highest-$Q$ rules fail while the ensemble selects the target; $32$ of these groups have $\sigma\ge0.09$. At the candidate level, the ensemble attains F1 $=0.761$, ROC--AUC $=0.980$, and MCC $=0.736$ at the validation-selected threshold.

The individual branches reach group-level accuracies of $0.823$ ($T_1$), $0.796$ ($T_2$), and $0.823$ ($N$). The stacked ensemble ($0.826$) exceeds the best single branch by one group, a difference that is not statistically significant (exact McNemar test, $p=1.0$); stacking therefore mainly combines the branches without loss of accuracy. To quantify the role of the field representation, both tree branches were retrained with the two-channel envelope descriptors replaced by descriptors of the modulus envelope alone (Table~\ref{tab:performance}, bottom). The group-level top-1 accuracy changed from $0.819$ to $0.823$ ($T_1$) and from $0.779$ to $0.796$ ($T_2$) in favor of the two-channel descriptors (exact McNemar test, $p=1.0$ and $p=0.46$), whereas the candidate-level precision decreased from $0.781$ to $0.757$ and from $0.649$ to $0.633$; the $95\%$ group-bootstrap intervals of these differences include zero. Within the resolution of the test set, separating the envelope into real- and imaginary-part channels therefore provides no measurable gain over the modulus envelope; the advantage of the learned selector over the frequency and quality-factor rules arises from combining the spatial field morphology with the resonance parameters and the group context.

\begin{table}[t]
\centering
\caption{Identification performance on the $299$ labeled groups of the chain test set: group-level top-1 accuracy by disorder range ($87$, $61$, and $151$ groups) and candidate-level metrics at the validation-selected threshold. The simple rules are given in Table~\ref{tab:simple_rules}. Bottom: ablation in which the tree branches use descriptors of the modulus envelope instead of the two-channel envelope; for these runs, only the overall group-level accuracy, F1, and ROC--AUC were recorded.}
\label{tab:performance}
\renewcommand{\arraystretch}{1.2}
\resizebox{\textwidth}{!}{%
\begin{tabular}{@{}lccccccc@{}}
\toprule
\textbf{Method} & \multicolumn{4}{c}{\textbf{Group-level top-1 accuracy}} & \textbf{F1} & \textbf{ROC--AUC} & \textbf{MCC} \\
\cmidrule(lr){2-5}
 & $\sigma\le0.05$ & $0.06$--$0.08$ & $\sigma\ge0.09$ & All $\sigma$ & & & \\
\midrule
LightGBM $T_1$ ($|\mathrm{Re}\,E|$, $|\mathrm{Im}\,E|$; group context) & 1.000 & 0.951 & 0.669 & 0.823 & 0.743 & 0.975 & 0.717 \\
LightGBM $T_2$ ($|\mathrm{Re}\,E|$, $|\mathrm{Im}\,E|$) & 1.000 & 0.820 & 0.669 & 0.796 & 0.665 & 0.963 & 0.629 \\
CNN--BiGRU $N$ & 1.000 & 0.902 & 0.689 & 0.823 & 0.745 & 0.978 & 0.719 \\
Stacked ensemble & 1.000 & 0.902 & 0.695 & 0.826 & 0.761 & 0.980 & 0.736 \\
\midrule
\multicolumn{8}{@{}l}{\textit{Ablation: modulus-envelope descriptors}} \\
LightGBM $T_1$ ($|E|$; group context) & --- & --- & --- & 0.819 & 0.739 & 0.975 & --- \\
LightGBM $T_2$ ($|E|$) & --- & --- & --- & 0.779 & 0.673 & 0.962 & --- \\
\bottomrule
\end{tabular}}
\end{table}

\subsection{$Q$ factor of the selected modes}
\label{subsec:chain-q}

The mean total $Q$ factor of the modes selected by the ensemble (Fig.~\ref{fig:figure3}(a)) follows that of the manually labeled modes at every disorder level and is consistent with the manually labeled reference data of Nazarov et al.~\cite{Nazarov2025_chain_disorder}. The model identifies the MQ mode even in realizations where several competing eigenmodes occur within a narrow spectral window; the direct evidence that it does not simply select the candidate with the highest $Q$ factor or the highest eigenfrequency is the set of $36$ test groups in which both rules fail and the ensemble is correct (Section~\ref{subsec:chain-performance}).

Because the total $Q$ contains all terms of Eq.~\eqref{eq:loss-budget}, its exponent is an effective one. Material absorption already limits the $Q$ factor of the ordered chain, $Q_0\approx8.9\times10^{3}$ (Section~\ref{subsec:ladder}), which flattens $\langle Q\rangle(\sigma)$ at small $\sigma$. The power-law fit in Fig.~\ref{fig:figure3}(a) is therefore restricted to $\sigma=0.03$--$0.15$, since at $\sigma=0.01$--$0.02$, $\langle Q\rangle$ stays within $10\%$ of $Q_0$. On the same $299$ labeled test groups, the fit gives $k=-0.80\pm0.05$ for the ensemble-selected modes and $k=-0.86\pm0.04$ for the manually labeled modes (standard errors of the fitted slopes); the difference is within the uncertainties. Figure~\ref{fig:figure3}(a) shows all $342$ test groups, i.e., it also includes the $43$ groups (all at $\sigma\ge0.06$) without an identified target, in which the ensemble still selects one candidate; with them, the ensemble gives $k=-0.84\pm0.04$. Over all $1{,}503$ labeled realizations of the dataset, the manual labels give $k=-0.86\pm0.03$. The sublinear decay of the total $Q$ is in line with the nearly linear decay reported for the MQ quasi-BIC of the same chain~\cite{Nazarov2025_chain_disorder}.

\subsection{Transfer to a phase-change-material platform}
\label{subsec:gst-transfer}

To assess the transferability of the workflow beyond the ceramic resonator chain, we considered a structurally and materially distinct platform based on phase-change-material resonators. The external dataset was computed for a periodic array of infinitely long Ge$_2$Sb$_2$Te$_5$ (GST) bars with rectangular cross-sections on a SiO$_2$ substrate (geometry as in Ref.~\cite{Vlasov2026GSTrobustness}). The lattice period is $a = 700$~nm, and the bar height and width are $h = 250$~nm and $w = 500$~nm, respectively~\cite{Vlasov2026GSTrobustness}. The system supports a dipole-like resonance and differs from the chain in its geometry, material, operating spectral range, and mode symmetry. Four datasets were considered: the amorphous and crystalline phases, each with lossless and lossy material models~\cite{Vlasov2026GSTrobustness}. Each dataset contains 10 values of $\sigma$ ($0.01$--$0.10$), $100$--$101$ realizations per value, and 20 modes per group (15 for the lossy crystalline phase). As in Ref.~\cite{Vlasov2026GSTrobustness}, disorder is introduced through independent fluctuations of the sidewall inclination angles of the bars, $\alpha_\ell=\alpha_0(1+R_\ell\sigma)$, where $\alpha_0=5^\circ$ and $R_\ell$ is uniformly distributed on $[-1,1]$. The GST envelopes were represented with the same number of samples as the chain envelopes ($1{,}883$); the models operate on the sample index and do not use the physical coordinate spacing.

\begin{figure}[h!]
\centering
\includegraphics[width=1\linewidth]{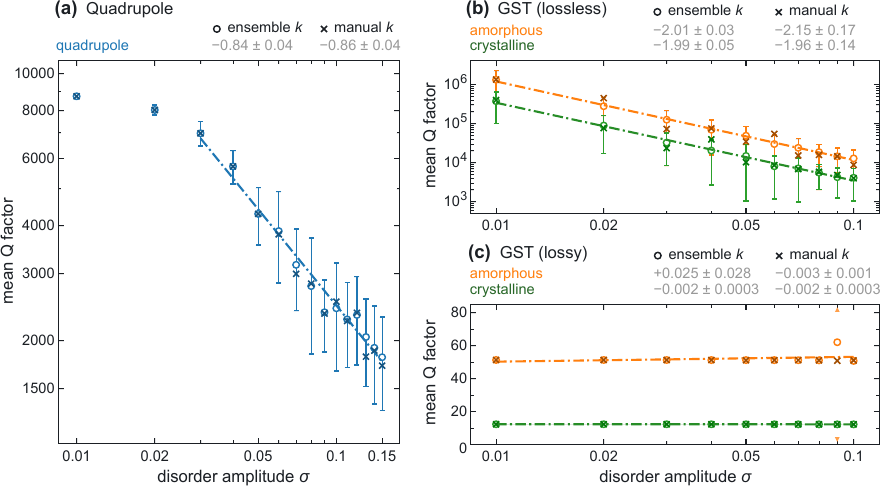}
\caption{Mean total $Q$ factor of the selected mode versus the disorder amplitude $\sigma$. (a)~MQ quasi-BIC of the ceramic-disk chain, test split. Open symbols: modes selected by the ensemble in all $342$ test groups; crosses: manually labeled modes ($299$ groups). The dash-dotted line is the power-law fit $\langle Q\rangle\propto\sigma^{k}$ for $\sigma=0.03$--$0.15$; the exponents ($\pm$ standard error) are given for the ensemble and for the manual labels (Table~\ref{tab:exponents_summary}). (b,~c)~GST bar metasurface without (b) and with (c) material losses, after adaptation (Section~\ref{subsec:gst-transfer}): amorphous (orange) and crystalline (green) phases; open symbols, ensemble selections in all groups (about $100$ per $\sigma$); crosses, the $50$ manually labeled groups ($5$ per $\sigma$); fits over $\sigma=0.01$--$0.10$. Error bars show the standard deviation over disorder realizations.}
\label{fig:figure3}
\end{figure}

In an ideal periodic structure, the collective mode extends across the resonator array and has a regular spatial energy distribution; structural disorder breaks this regularity and can induce realization-specific localization patterns, spectral shifts, and changes in radiative losses. Applied without adaptation, the ensemble trained only on the chain selects the target in $0\%$, $4\%$, $0\%$, and $0\%$ of the labeled groups (amorphous lossless, crystalline lossless, amorphous lossy, and crystalline lossy GST, respectively), so the model must be adapted to the new platform. The only labeled GST data are 5 realizations for each of the 10 values of $\sigma$, i.e., 50 groups per dataset ($5\%$ of the data). Adaptation was therefore evaluated by grouped five-fold cross-validation over these 50 groups: in each fold, 32 groups were used for updating the models, 8 for early stopping and stacking, and 10 for held-out evaluation, so that every labeled group is evaluated exactly once by models that have not seen it. Both tree branches continue boosting from the chain-trained models for up to $500$ additional iterations, with early stopping after $20$ iterations without an improvement of the log-loss on the early-stopping groups. The neural branch is fine-tuned from the chain-trained weights with the objective of Eq.~\eqref{eq:hybrid-loss} (Adam, learning rate $10^{-4}$, at most $100$ epochs), and the epoch with the lowest ranking loss on the early-stopping groups is retained. The stacking coefficients are refitted on the early-stopping groups with an $L_2$ penalty toward the chain-trained coefficients. All three seeds are adapted, and their probabilities are averaged. The unlabeled groups (about $95\%$ of each dataset) are scored by the average of the five fold models, and the quality-factor curves in Fig.~\ref{fig:figure3}(b,c) include all groups.

After adaptation, the ensemble selects the target mode in $50/50$ (amorphous, lossless), $49/50$ (crystalline, lossless), $50/50$ (amorphous, lossy), and $49/50$ (crystalline, lossy) of the labeled groups (Fig.~\ref{fig:figure4}(c)). In every dataset, one of the simple rules already reaches $0.94$--$1.00$: in the lossless datasets, the target is the highest-$Q$ candidate (except in one crystalline group), whereas in the lossy datasets the highest-$Q$ candidate is never the target ($0/50$) and the frequency rule (the candidate at a fixed frequency rank, fitted on the other labeled groups) reaches $1.00$ (amorphous) and $0.94$ (crystalline). The adapted ensemble matches the best rule in three datasets and exceeds it for lossy crystalline GST ($0.98$ versus $0.94$), where five candidates of each group have $Q$ factors within $10\%$ of that of the target, so that $Q$ carries no information about the mode. In all $1{,}010$ groups of this dataset, the selected mode has $Q=12.05$--$12.23$. Its single error on lossless crystalline GST occurs in the only group of that dataset in which the target is not the highest-$Q$ candidate. In the amorphous lossy dataset, all but 4 of the $1{,}008$ groups select a mode of the target family ($Q\approx51$); the four exceptions are unlabeled groups at $\sigma=0.09$ and $0.10$, and one of them, a non-target mode with $Q\approx1.2\times10^{3}$, produces the outlier at $\sigma=0.09$ in Fig.~\ref{fig:figure3}(c).

Table~\ref{tab:exponents_summary} summarizes the fitted exponents. For the lossless GST datasets, the exponents of the modes selected by the adapted ensemble, $-2.01\pm0.03$ (amorphous) and $-1.99\pm0.05$ (crystalline), agree with the inverse-quadratic law derived analytically and confirmed numerically for random sidewall-angle fluctuations in this structure~\cite{Vlasov2026GSTrobustness}; the 50 manually labeled groups give $-2.15\pm0.17$ and $-1.96\pm0.14$. For the lossy datasets, the total $Q$ of the target mode is nearly independent of $\sigma$ ($\approx12$ for the crystalline and $\approx51$ for the amorphous phase), reflecting absorption-dominated losses~\cite{Vlasov2026GSTrobustness}; the fitted exponents are $+0.025\pm0.028$ and $-0.0020\pm0.0003$. In this case, the total $Q$ is not a measure of the disorder-induced losses, and errors in the selection appear as outliers in $\langle Q\rangle$, such as the single amorphous group at $\sigma=0.09$ discussed above. These results show that, after adaptation on a small fraction of target-domain data, the workflow can be transferred to a platform with a different geometry, material, and mode symmetry.

\begin{table}[t]
\centering
\caption{Fitted exponents $k$ of the mean total $Q$ factor, $\langle Q\rangle\propto\sigma^{k}$ ($\pm$ standard error of the slope), for the manually labeled modes and for the modes selected by the ensemble, as in Fig.~\ref{fig:figure3}. Fit windows: $\sigma=0.03$--$0.15$ (chain) and $0.01$--$0.10$ (GST). Chain: manual labels of the $299$ labeled test groups; ensemble, all $342$ test groups ($-0.80\pm0.05$ on the $299$ labeled groups). GST: manual labels of the $50$ labeled groups per dataset; ensemble, all groups (about $1{,}000$ per dataset). Analytical value: inverse-quadratic scaling for sidewall-angle disorder without material loss~\cite{Vlasov2026GSTrobustness}.}
\label{tab:exponents_summary}
\renewcommand{\arraystretch}{1.2}
\setlength{\tabcolsep}{4pt}
\begin{tabularx}{\textwidth}{@{}l X X c@{}}
\toprule
\textbf{Dataset} & \textbf{Manual labels} & \textbf{Ensemble} & \textbf{Analytics} \\
\midrule
MQ ceramic chain & $-0.86\pm0.04$ & $-0.84\pm0.04$ & --- \\
Amorphous GST, lossless & $-2.15\pm0.17$ & $-2.01\pm0.03$ & $-2$ \\
Crystalline GST, lossless & $-1.96\pm0.14$ & $-1.99\pm0.05$ & $-2$ \\
Amorphous GST, lossy & $-0.003\pm0.001$ & $+0.025\pm0.028$ & --- \\
Crystalline GST, lossy & $-0.0017\pm0.0003$ & $-0.0020\pm0.0003$ & --- \\
\bottomrule
\end{tabularx}
\end{table}

\subsection{Limitations}
\label{subsec:limitations}

The present analysis covers $\sigma\le0.15$ and a single chain geometry. The selector always returns one candidate per group; groups without an unambiguously identified target ($204$ in the dataset, $43$ in the test split) are excluded from the group-level top-1 evaluation, and a calibrated rejection mechanism remains to be developed. Using the candidate-level threshold $t$ as a rejection rule ($\max_i p_i<t$) would reject $27$ of these $43$ test groups, but also $57$ of the $299$ groups with an identified target. The unsigned component envelopes depend on the arbitrary global phase of the eigenvector, and their invariance to a global phase rotation has not been established; the modulus envelope, which does not depend on the phase, performs equally well (Section~\ref{subsec:chain-performance}). The $\mathrm{norm}_{\max}$-derived and group-size descriptors make the complete predictor dependent on the eigensolver normalization convention and settings; in particular, the GST groups (15--20 modes) lie outside the range of the chain groups (8--12 modes). For GST, the only labeled data are the 50 groups per dataset used for adaptation and its cross-validated evaluation, and the benefit of chain pretraining relative to training the same architecture from scratch on the same target-domain groups remains to be quantified. At large $\sigma$, errors in the selection appear as outliers in the mean $Q$ factor.

\section{Conclusions}

Disorder reorders the ladder of standing waves at the edge of the BIC band: the highest-frequency mode remains the target up to $\sigma\approx0.06$, whereas the highest-$Q$ mode fails from $\sigma\approx0.04$. Above $\sigma\approx0.07$, modes localized on a few disks appear above the band edge, and neither frequency nor $Q$ factor identifies the target. In this regime ($\sigma\ge0.09$), the ensemble that combines the spatial field envelope with the resonance parameters identifies the target in $69.5\%$ of realizations, compared with at most $57.6\%$ for the frequency- and $Q$-based rules considered; over all held-out realizations, it reaches $82.6\%$. The mean total $Q$ factor of the selected modes reproduces that of the manually labeled modes ($k=-0.80\pm0.05$ versus $-0.86\pm0.04$ on the same realizations). Adaptation on 50 labeled realizations per dataset transfers the workflow to a Ge$_2$Sb$_2$Te$_5$ metasurface ($98$--$100\%$ of the labeled realizations, evaluated out of fold), including the lossy crystalline case, where the $Q$ factor carries no information about the mode. The workflow provides a reproducible and scalable post-processing tool for multimode eigenmode datasets of disordered photonic structures.

\section*{Acknowledgments}

The work is funded by Russian Science Foundation grant No 26-72-10063. E.M. is grateful to the Priority 2030 Federal Academic Leadership Program. 

\bibliographystyle{elsarticle-num}
\bibliography{references}

@article{Hsu2016_BIC_review,
  author = {Hsu, Chia Wei and Zhen, Bo and Stone, A. Douglas and Joannopoulos, John D. and Soljačić, Marin},
  title = {Bound states in the continuum},
  journal = {Nature Reviews Materials},
  volume = {1},
  number = {9},
  pages = {16048},
  year = {2016},
  doi = {10.1038/natrevmats.2016.48},
}

@article{Koshelev2018_asymmetricBIC,
  author = {Koshelev, Kirill and Lepeshov, Sergey and Liu, Mingkai and Bogdanov, Andrey and Kivshar, Yuri},
  title = {Asymmetric Metasurfaces with High-{$Q$} Resonances Governed by Bound States in the Continuum},
  journal = {Physical Review Letters},
  volume = {121},
  number = {19},
  pages = {193903},
  year = {2018},
  doi = {10.1103/physrevlett.121.193903},
}

@article{Kodigala2017_laser,
  author = {Kodigala, Ashok and Lepetit, Thomas and Gu, Qing and Bahari, Babak and Fainman, Yeshaiahu and Kanté, Boubacar},
  title = {Lasing action from photonic bound states in continuum},
  journal = {Nature},
  volume = {541},
  number = {7636},
  pages = {196--199},
  year = {2017},
  doi = {10.1038/nature20799},
}

@article{Sidorenko2021_chainBIC,
  author = {Sidorenko, Mikhail S. and Sergaeva, Olga N. and Sadrieva, Zarina F. and Roques-Carmes, Charles and Muraev, Pavel S. and Maksimov, Denis N. and Bogdanov, Andrey A.},
  title = {Observation of an Accidental Bound State in the Continuum in a Chain of Dielectric Disks},
  journal = {Physical Review Applied},
  volume = {15},
  number = {3},
  pages = {034041},
  year = {2021},
  doi = {10.1103/physrevapplied.15.034041},
}

@article{Sadrieva2019_chainBIC,
  author = {Sadrieva, Zarina F. and Belyakov, Mikhail A. and Balezin, Mikhail A. and Kapitanova, Polina V. and Nenasheva, Elena A. and Sadreev, Almas F. and Bogdanov, Andrey A.},
  title = {Experimental observation of a symmetry-protected bound state in the continuum in a chain of dielectric disks},
  journal = {Physical Review A},
  volume = {99},
  number = {5},
  pages = {053804},
  year = {2019},
  doi = {10.1103/physreva.99.053804},
}

@article{Nazarov2025_chain_disorder,
  author = {Nazarov, Ravshanjon and Khanabiev, Denis and Chernysheva, Elizaveta and Dudnikova, Alexandra and Istomin, Vyacheslav and Sidorenko, Mikhail and Shi, Jinhui and Maslova, Ekaterina and Bogdanov, Andrey and Kondratenko (Sadrieva), Zarina},
  title = {Bound states in the continuum in a chain of coupled {Mie} resonators with structural disorder: theory and experiment},
  journal = {Nanophotonics},
  volume = {14},
  number = {19},
  pages = {3133--3143},
  year = {2025},
  doi = {10.1515/nanoph-2025-0225},
}

@article{ML_photonics_review,
  author = {Chen, Yuheng and Montes McNeil, Alexander and Park, Taehyuk and Wilson, Blake A. and Iyer, Vaishnavi and Bezick, Michael and Choi, Jae-Ik and Ojha, Rohan and Mahendran, Pravin and Singh, Daksh Kumar and Chitturi, Geetika and Chen, Peigang and Do, Trang and Kildishev, Alexander V. and Shalaev, Vladimir M. and Moebius, Michael and Cai, Wenshan and Liu, Yongmin and Boltasseva, Alexandra},
  title = {Machine-learning-assisted photonic device development: a multiscale approach from theory to characterization},
  journal = {Nanophotonics},
  volume = {14},
  number = {23},
  pages = {3761--3793},
  year = {2025},
  doi = {10.1515/nanoph-2025-0049},
}

@article{PhysicsInformedModes,
  author = {Ghosh, Abantika and Elhamod, Mohannad and Bu, Jie and Lee, Wei-Cheng and Karpatne, Anuj and Podolskiy, Viktor A},
  title = {Physics-Informed Machine Learning for Optical Modes in Composites},
  journal = {Advanced Photonics Research},
  volume = {3},
  number = {11},
  pages = {2200073},
  year = {2022},
  doi = {10.1002/adpr.202200073},
}

@article{Jorkowski2019modetracking,
  author = {Jorkowski, Philipp and Schuhmann, Rolf},
  title = {Mode tracking for parametrized eigenvalue problems in computational electromagnetics},
  journal = {Applied Computational Electromagnetics Society (ACES) Journal},
  volume = {34},
  number = {2},
  pages = {252--257},
  year = {2019},
}

@misc{Vlasov2026GSTrobustness,
  author = {Vlasov, Nikolai A. and Solomonov, Alexander I. and Kondratenko, Zarina F. and Bronnikov, Kirill A. and Rybin, Mikhail V. and Maslova, Ekaterina E.},
  title = {Robustness of bound states in the continuum in metasurface based on {Ge}{$_2$}{Sb}{$_2$}{Te}{$_5$} versus structural imperfections},
  year = {2026},
  doi = {10.48550/arXiv.2605.05779},
  eprint = {2605.05779},
  archiveprefix = {arXiv},
}

@article{Koshelev2023-review,
  author = {Koshelev, Kirill L. and Sadrieva, Zarina F. and Shcherbakov, Alexey A. and Kivshar, Yu.S. and Bogdanov, Andrey A.},
  title = {Bound states in the continuum in photonic structures},
  journal = {Physics-Uspekhi},
  volume = {66},
  number = {05},
  pages = {494--517},
  year = {2023},
  doi = {10.3367/ufne.2021.12.039120},
}

@article{Zhang2026-review,
  author = {Zhang, Shubin and Fan, Ye and Ma, Yufei and Zong, Meixue and Sun, Handong and Yang, Zhiqiang and Xu, Zhengji},
  title = {Bound states in the continuum: From fundamental physics to emerging photonic paradigms},
  journal = {iScience},
  volume = {29},
  number = {2},
  pages = {114803},
  year = {2026},
  doi = {10.1016/j.isci.2026.114803},
}

@article{Kang2023-applications,
  author = {Kang, Meng and Liu, Tao and Chan, C. T. and Xiao, Meng},
  title = {Applications of bound states in the continuum in photonics},
  journal = {Nature Reviews Physics},
  volume = {5},
  number = {11},
  pages = {659--678},
  year = {2023},
  doi = {10.1038/s42254-023-00642-8},
}

@article{Bulgakov2019-finite,
  author = {Bulgakov, Evgeny N. and Sadreev, Almas F.},
  title = {High-{$Q$} resonant modes in a finite array of dielectric particles},
  journal = {Physical Review A},
  volume = {99},
  number = {3},
  pages = {033851},
  year = {2019},
  doi = {10.1103/physreva.99.033851},
}

@article{Sadrieva2017-roughness,
  author = {Sadrieva, Zarina F. and Sinev, Ivan S. and Koshelev, Kirill L. and Samusev, Anton and Iorsh, Ivan V. and Takayama, Osamu and Malureanu, Radu and Bogdanov, Andrey A. and Lavrinenko, Andrei V.},
  title = {Transition from Optical Bound States in the Continuum to Leaky Resonances: Role of Substrate and Roughness},
  journal = {ACS Photonics},
  volume = {4},
  number = {4},
  pages = {723--727},
  year = {2017},
  doi = {10.1021/acsphotonics.6b00860},
}

@article{Ni2017-fluctuations,
  author = {Ni, Liangfu and Jin, Jicheng and Peng, Chao and Li, Zhengbin},
  title = {Analytical and statistical investigation on structural fluctuations induced radiation in photonic crystal slabs},
  journal = {Optics Express},
  volume = {25},
  number = {5},
  pages = {5580},
  year = {2017},
  doi = {10.1364/oe.25.005580},
}

@article{Centeno2025-emission,
  author = {Centeno, Emmanuel and Kalt, Victor and Smaali, Rafik and Moreau, Antoine},
  title = {Disorder-induced effects on spontaneous emission from quasibound states in the continuum in regular and dimer gratings},
  journal = {Physical Review B},
  volume = {112},
  number = {13},
  pages = {134202},
  year = {2025},
  doi = {10.1103/cmdf-jdg1},
}

@article{Maslova2021-disorder,
  author = {Maslova, Ekaterina E. and Rybin, Mikhail V. and Bogdanov, Andrey A. and Sadrieva, Zarina F.},
  title = {Bound states in the continuum in periodic structures with structural disorder},
  journal = {Nanophotonics},
  volume = {10},
  number = {17},
  pages = {4313--4321},
  year = {2021},
  doi = {10.1515/nanoph-2021-0475},
}

@article{Segev2013-anderson,
  author = {Segev, Mordechai and Silberberg, Yaron and Christodoulides, Demetrios N.},
  title = {{Anderson} localization of light},
  journal = {Nature Photonics},
  volume = {7},
  number = {3},
  pages = {197--204},
  year = {2013},
  doi = {10.1038/nphoton.2013.30},
}

@article{Lahini2008-anderson,
  author = {Lahini, Yoav and Avidan, Assaf and Pozzi, Francesca and Sorel, Marc and Morandotti, Roberto and Christodoulides, Demetrios N. and Silberberg, Yaron},
  title = {{Anderson} Localization and Nonlinearity in One-Dimensional Disordered Photonic Lattices},
  journal = {Physical Review Letters},
  volume = {100},
  number = {1},
  pages = {013906},
  year = {2008},
  doi = {10.1103/physrevlett.100.013906},
}

@article{Poddubny2012-fano,
  author = {Poddubny, Alexander N. and Rybin, Mikhail V. and Limonov, Mikhail F. and Kivshar, Yuri S.},
  title = {{Fano} interference governs wave transport in disordered systems},
  journal = {Nature Communications},
  volume = {3},
  number = {1},
  pages = {914},
  year = {2012},
  doi = {10.1038/ncomms1924},
}

@article{Minkov2013-statistics,
  author = {Minkov, Momchil and Dharanipathy, Ulagalandha Perumal and Houdré, Romuald and Savona, Vincenzo},
  title = {Statistics of the disorder-induced losses of high-{Q} photonic crystal cavities},
  journal = {Optics Express},
  volume = {21},
  number = {23},
  pages = {28233},
  year = {2013},
  doi = {10.1364/oe.21.028233},
}

@article{Ma2021-deep,
  author = {Ma, Wei and Liu, Zhaocheng and Kudyshev, Zhaxylyk A. and Boltasseva, Alexandra and Cai, Wenshan and Liu, Yongmin},
  title = {Deep learning for the design of photonic structures},
  journal = {Nature Photonics},
  volume = {15},
  number = {2},
  pages = {77--90},
  year = {2021},
  doi = {10.1038/s41566-020-0685-y},
}

@article{Jiang2021-deepnn,
  author = {Jiang, Jiaqi and Chen, Mingkun and Fan, Jonathan A.},
  title = {Deep neural networks for the evaluation and design of photonic devices},
  journal = {Nature Reviews Materials},
  volume = {6},
  number = {8},
  pages = {679--700},
  year = {2021},
  doi = {10.1038/s41578-020-00260-1},
}

@article{Peurifoy2018-nanophotonic,
  author = {Peurifoy, John and Shen, Yichen and Jing, Li and Yang, Yi and Cano-Renteria, Fidel and DeLacy, Brendan G. and Joannopoulos, John D. and Tegmark, Max and Soljačić, Marin},
  title = {Nanophotonic particle simulation and inverse design using artificial neural networks},
  journal = {Science Advances},
  volume = {4},
  number = {6},
  pages = {eaar4206},
  year = {2018},
  doi = {10.1126/sciadv.aar4206},
}

@inproceedings{Li2026-metadit,
  author = {Li, Hao and Bogdanov, Andrey},
  title = {{MetaDiT}: Enabling Fine-grained Constraints in High-degree-of Freedom Metasurface Design},
  booktitle = {Proceedings of the AAAI Conference on Artificial Intelligence},
  volume = {40},
  number = {1},
  pages = {606--614},
  year = {2026},
  doi = {10.1609/aaai.v40i1.37025},
}

@article{Christensen2020-crystals,
  author = {Christensen, Thomas and Loh, Charlotte and Picek, Stjepan and Jakobović, Domagoj and Jing, Li and Fisher, Sophie and Ceperic, Vladimir and Joannopoulos, John D. and Soljačić, Marin},
  title = {Predictive and generative machine learning models for photonic crystals},
  journal = {Nanophotonics},
  volume = {9},
  number = {13},
  pages = {4183--4192},
  year = {2020},
  doi = {10.1515/nanoph-2020-0197},
}

@article{Barth2018-modes,
  author = {Barth, Carlo and Becker, Christiane},
  title = {Machine learning classification for field distributions of photonic modes},
  journal = {Communications Physics},
  volume = {1},
  number = {1},
  pages = {58},
  year = {2018},
  doi = {10.1038/s42005-018-0060-1},
}

@article{Molokeev2023-forest,
  author = {Molokeev, M. S. and Kostyukov, A. S. and Ershov, A. E. and Maksimov, D. N. and Gerasimov, V. S. and Polyutov, S. P.},
  title = {Infrared bound states in the continuum: random forest method},
  journal = {Optics Letters},
  volume = {48},
  number = {17},
  pages = {4460},
  year = {2023},
  doi = {10.1364/OL.494629},
}

@article{Ma2022-bic,
  author = {Ma, Xuezhi and Ma, Yuan and Cunha, Preston and Liu, Qiushi and Kudtarkar, Kaushik and Xu, Da and Wang, Jiafei and Chen, Yixin and Wong, Zi Jing and Liu, Ming and Hipwell, M. Cynthia and Lan, Shoufeng},
  title = {Strategical Deep Learning for Photonic Bound States in the Continuum},
  journal = {Laser \& Photonics Reviews},
  volume = {16},
  number = {10},
  pages = {2100658},
  year = {2022},
  doi = {10.1002/lpor.202100658},
}

@article{Raines2012-CMtracking,
  author = {Raines, Bryan D. and Rojas, Roberto G.},
  title = {Wideband characteristic mode tracking},
  journal = {IEEE Transactions on Antennas and Propagation},
  volume = {60},
  number = {7},
  pages = {3537--3541},
  year = {2012},
  doi = {10.1109/TAP.2012.2196914},
}

@article{Georg2019-UQtracking,
  author = {Georg, Niklas and Ackermann, Wolfgang and Corno, Jacopo and Sch{\"o}ps, Sebastian},
  title = {Uncertainty quantification for {Maxwell's} eigenproblem based on isogeometric analysis and mode tracking},
  journal = {Computer Methods in Applied Mechanics and Engineering},
  volume = {350},
  pages = {228--244},
  year = {2019},
  doi = {10.1016/j.cma.2019.03.002},
}

@article{Ziegler2023-morphing,
  author = {Ziegler, Anna and Georg, Niklas and Ackermann, Wolfgang and Sch{\"o}ps, Sebastian},
  title = {Mode recognition by shape morphing for {Maxwell's} eigenvalue problem in cavities},
  journal = {IEEE Transactions on Antennas and Propagation},
  volume = {71},
  number = {5},
  pages = {4315--4325},
  year = {2023},
  doi = {10.1109/TAP.2023.3249907},
}

@article{Li2026-MLCMtracking,
  author = {Li, Hui and Li, Zhaolin and Qu, Chang and Zhang, Jiapeng},
  title = {Machine learning assisted approach for characteristic mode tracking},
  journal = {Microwave and Optical Technology Letters},
  volume = {68},
  number = {10},
  pages = {e70811},
  year = {2026},
  doi = {10.1002/mop.70811},
}

@article{Wuttig2017-pcm,
  author = {Wuttig, M. and Bhaskaran, H. and Taubner, T.},
  title = {Phase-change materials for non-volatile photonic applications},
  journal = {Nature Photonics},
  volume = {11},
  number = {8},
  pages = {465--476},
  year = {2017},
  doi = {10.1038/nphoton.2017.126},
}

@article{Bochek2022-gst,
  author = {Bochek, Daria V. and Solodovchenko, Nikolay S. and Yavsin, Denis A. and Pevtsov, Alexander B. and Samusev, Kirill B. and Limonov, Mikhail F.},
  title = {Bound states in the continuum versus material losses: {Ge}{$_2$}{Sb}{$_2$}{Te}{$_5$} as an example},
  journal = {Physical Review B},
  volume = {105},
  number = {16},
  pages = {165425},
  year = {2022},
  doi = {10.1103/physrevb.105.165425},
}

@inproceedings{Bahdanau2015-attention,
  author = {Bahdanau, Dzmitry and Cho, Kyunghyun and Bengio, Yoshua},
  title = {Neural Machine Translation by Jointly Learning to Align and Translate},
  booktitle = {3rd International Conference on Learning Representations (ICLR)},
  year = {2015},
  eprint = {1409.0473},
  archiveprefix = {arXiv},
}

@inproceedings{Yang2016-HAN,
  author = {Yang, Zichao and Yang, Diyi and Dyer, Chris and He, Xiaodong and Smola, Alex and Hovy, Eduard},
  title = {Hierarchical Attention Networks for Document Classification},
  booktitle = {Proceedings of the 2016 Conference of the North American Chapter of the Association for Computational Linguistics: Human Language Technologies},
  pages = {1480--1489},
  year = {2016},
  doi = {10.18653/v1/N16-1174},
}

@article{Harsh2026-hybridmode,
  author = {Harsh, D. R. and Khatri, Narender and Awasthi, Suneet Kumar},
  title = {Hybrid machine learning approach for mode classification of {SPR} driven photonic crystal fiber sensor},
  journal = {Results in Physics},
  volume = {88},
  pages = {108746},
  year = {2026},
  doi = {10.1016/j.rinp.2026.108746},
}

@article{Han2023-scardetector,
  author = {Han, Chen-Di and Wang, Cheng-Zhen and Lai, Ying-Cheng},
  title = {Meta-Machine-Learning-Based Quantum Scar Detector},
  journal = {Physical Review Applied},
  volume = {19},
  number = {6},
  pages = {064042},
  year = {2023},
  doi = {10.1103/physrevapplied.19.064042},
}

@article{Liu2026-cmtnn,
  author = {Liu, Song-Yi and Zhong, Hao-Tian and Yu, Xiao-Chong and Zhang, Bo-Lun and Zhang, Liu-Yang and Xu, Yi and Zhu, Ning-Hua and Chen, Jin-hui and Wen, Hua-Shun and Huang, Shan-Guo and Yang, Da-Quan},
  title = {Deciphering optical coupled resonant systems with physics-data co-driven deep neural networks},
  journal = {Light: Science \& Applications},
  volume = {15},
  number = {1},
  pages = {279},
  year = {2026},
  doi = {10.1038/s41377-026-02389-0},
}

@article{Laude2026-boundary,
  author = {Laude, Vincent and Röhlig, David},
  title = {Identifying bound states in the continuum by their boundary sensitivity},
  journal = {Physical Review Applied},
  volume = {25},
  number = {4},
  pages = {044086},
  year = {2026},
  doi = {10.1103/998l-j9cc},
}

@article{Cybinski2025-ood,
  author = {Cybiński, Kacper and Płodzień, Marcin and Tomza, Michał and Lewenstein, Maciej and Dauphin, Alexandre and Dawid, Anna},
  title = {Characterizing out-of-distribution generalization of neural networks: application to the disordered {Su}--{Schrieffer}--{Heeger} model},
  journal = {Machine Learning: Science and Technology},
  volume = {6},
  number = {1},
  pages = {015014},
  year = {2025},
  doi = {10.1088/2632-2153/ad9079},
}

\end{document}